\documentclass[prb,superscriptaddress,twocolumn]{revtex4-1}

\usepackage{color}
\usepackage{graphicx}
\usepackage{amssymb}
\usepackage[hidelinks=true]{hyperref}
\hypersetup{
	colorlinks = true,
	urlcolor = blue,
	linkcolor = blue,
	citecolor = blue
}

\begin{document}

\title{Interface enhanced magnetic anisotropy in Pt/EuO$_{1-x}$ films}
\date{\today}
\author{Gaurab Rimal}
\altaffiliation{Present address: Department of Physics and Astronomy, Rutgers University, Piscataway, NJ 08854, USA}
\author{Jinke Tang}
\altaffiliation{Corresponding author: jtang2@uwyo.edu}
\affiliation{Department of Physics \& Astronomy, University of Wyoming, Laramie, WY 82071, USA}

\begin{abstract}
	We report proximity effects of spin-orbit coupling in EuO$_{1-x}$ films capped with a Pt overlayer.  Transport measurements suggest that current flows along a conducting channel at the interface between the Pt and EuO.  The temperature dependence of the resistivity picks up the critical behaviors of EuO, i.e., the metal-to-insulator transition. We also find an unusual enhancement of the magnetic anisotropy in this structure from its bulk value which results from strong spin-orbit coupling across the Pt/EuO$_{1-x}$ interface.  
\end{abstract}

\maketitle

%%%% macros
\newcommand{\red}[1]{\textcolor{red}{#1}}
\newcommand{\blue}[1]{\textcolor{blue}{#1}}
\newcommand{\green}[1]{\textcolor{green}{#1}}

\section{Introduction}
At the interface between two different materials, symmetry breaking and spin-orbit coupling can lead to unusual and exciting physics \cite{Hellman2017,Mannhart2010}.  The properties of the interface can be drastically different than the constituent layers, as evidenced by the formation of 2D electron gas (2DEG) at the interface of various semiconductor and insulator systems.  The most surprising discovery was the observation of 2DEG at the interface of SrTiO$_3$ and LaAlO$_3$ \cite{Ohtomo2004}, which are both insulators.  These interface 2DEG have been found to host many exotic states which lead to effects such as the quantum Hall effects, superconductivity \cite{Reyren2007} and the quantum spin Hall effect \cite{Konig2007}.

In magnetic systems, normal metal - magnetic insulator interface has been widely investigated for possible applications to spintronics.  Effects such as current induced magnetization switching \cite{Avci2016,Li2016} and the spin Hall magnetoresistance \cite{Nakayama2013,Lu2013} have highlighted the importance of spin-orbit coupling at the interface of a magnetic insulator and a normal metal. In addition, much emphasis has been placed in the discovery and applications of magnetic semiconductors for spintronics. 

Europium monoxide (EuO) is a well-studied ferromagnetic semiconductor. After its discovery \cite{Matthias1961}, the study of the material has shown many interesting properties.  Stoichiometric EuO has ferromagnetic transition at 69 K, but electron doping enhances the Curie temperature up to around 150 K at ambient pressure.  This enhancement is a result of added charge carriers, and various mechanisms have been invoked to study its properties \cite{Mauger1977,Arnold2008,Stollenwerk2015}.  A recent first-principles study found that the competition between direct and indirect exchange couplings dictates many of the properties of electron doped EuO \cite{An2013}.  Although still debated, most studies point the mechanism of Curie temperature enhancement to bound magnetic polarons \cite{Kasuya1968,Liu2013,Torrance1972}, although other models such as RKKY and the Kondo lattice model are also invoked to explain the various properties observed for EuO \cite{Schiller2001,Sinjukow2003}.

Although bulk EuO is topologically trivial, theoretical studies have found that the interface between EuO and other materials may host topological states.  For example, EuO/CdO superlattice was found to host Weyl nodes and quantum anomalous Hall states \cite{Zhang2014}.  It was also found that the interface of magnetic insulators and heavy metals \cite{Garrity2013} and the EuO/GdN interface \cite{Garrity2014} may be Chern insulators.  There has also been an experimental observation of topological Hall effect in EuO films, suggesting the formation of 2D skyrmions \cite{Ohuchi2015}.  Due to their large magnetization (7 $\mu_B$/Eu) as well as highly insulating behavior, EuO and EuS have also been used to study magnetic proximity effects.  Examples include the large magnetic exchange field induced in graphene \cite{Wei2016} as well as the room temperature ferromagnetism observed at Bi$_2$Se$_3$ interface as a result of large spin-orbit coupling and spin-momentum locking \cite{Katmis2016}. Thus EuO is an exciting system for the exploration of new physics.

Here, we report on the transport properties at the Pt/EuO interface and show unusual behaviors including enhanced magnetic anisotropy at the interface, which results from the strong spin-orbit coupling and symmetry breaking.  This result is consistent with recent calculations that show strong interfacial Eu(4f)- Pt(5d) coupling \cite{Fredrickson2016} and hints at the possibility for the formation of spin-polarized localized mid-gap electronic state in the first two atomic layers of EuO next to the Pt.

\section{Experiments and results}

\begin{figure}[ht]
	\centering
	\includegraphics[width=0.9\linewidth]{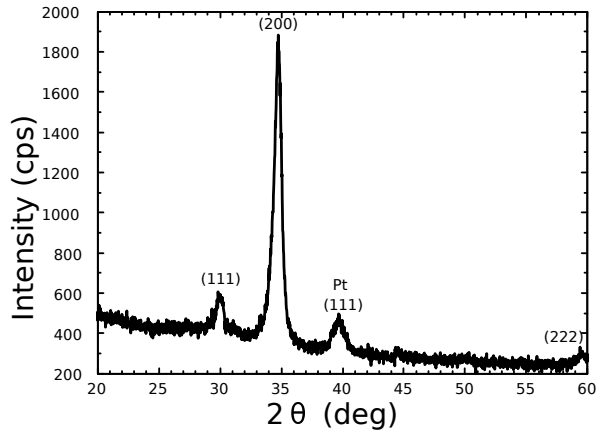}
	\caption{X-ray diffraction pattern for Pt/EuO film.}
	\label{fig:fig0}
\end{figure}

We deposited EuO$_{1-x}$ (hereafter simply EuO) on a Si (001) substrate using pulsed laser deposition (PLD).  The Eu metal target was kept about 7 cm from the substrate, and was ablated using a Nd:YAG laser with 266 nm wavelength and energy of 320 mJ/pulse.  The substrate was heated at a temperature of 350$^\circ$C and the base pressure was better than $2\times10^{-6}$ Pa.  Since it is easy to form Eu$_2$O$_3$, we refrained from using oxygen and instead relied on residual oxygen in the chamber.  We did not etch the substrate, because of two reasons: first, the native oxide at the surface of the substrate would prevent the reaction between Eu and Si and, second, provide oxygen to Eu to form the EuO seed layer. Both these functions are facilitated by keeping the SiO$_2$ layer when we evaluate the interface free energy \cite{Caspers2013_dissertation}.  This technique is comparable to the ``distillation method" that has been used to deposit high quality EuO using molecular beam epitaxy (MBE) \cite{Sutarto2009}.  A thin layer of Pt ($\sim$20 nm) was then deposited in-situ at the same temperature, also using PLD. The growth rate was about 0.2 nm/s and it was cooled at 10 $^\circ$C/min after the deposition of Pt.  We were able to grow phase pure EuO using this technique, which was verified using x-ray diffraction (XRD) as shown in Fig. \ref{fig:fig0}.  The thickness of the entire film was around 600 nm.

\begin{figure}[ht]
	\centering
	\includegraphics[width=0.9\linewidth]{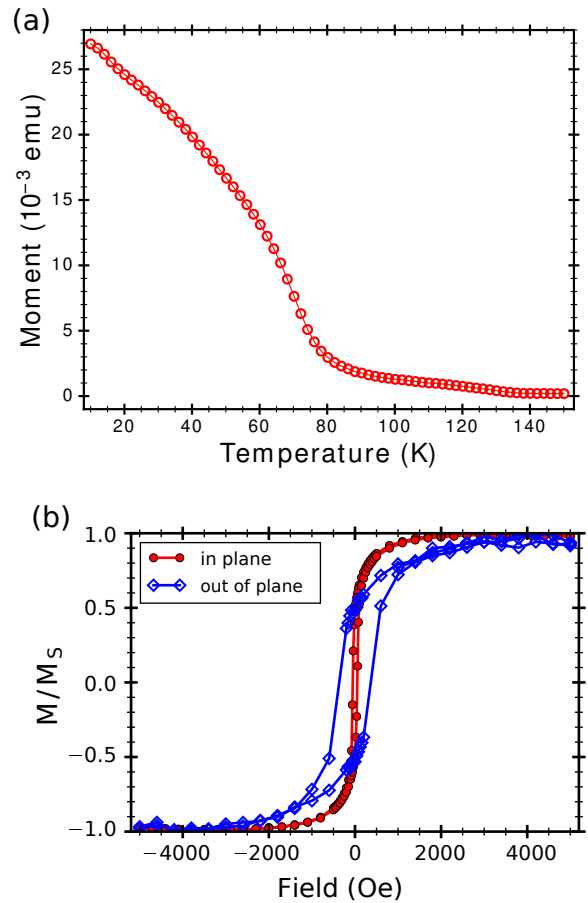}
	\caption{(a) The temperature dependent magnetic moment for the Pt/EuO film. (b) The magnetic hysteresis curves at 5K in different orientations. The curves are normalized to their saturation values.}
	\label{fig:fig1}
\end{figure}

It is likely that our preparation technique makes the films oxygen deficient, and is thus electron doped.  To test this we performed both AC and DC magnetometry to determine the magnetic properties of the film.  Indeed, we clearly observe a transition around 130 K, characteristic of oxygen deficient samples, as seen in Fig. \ref{fig:fig1}(a). We also observe an in-plane magnetic anisotropy, which is shown in  Fig. \ref{fig:fig1}(b). The saturation moment is approximated to about 6 $\mu_B$/Eu, which is close to the expected value of 7 $\mu_B$.  Since our films are polycrystalline, the cause of this anisotropy is most likely the film shape.  The coercivity for the in-plane direction is 52 Oe while the out-of-plane coercivity is 367 Oe at 5 K.  Some other factors influencing the anisotropy also include the interface and interdiffusion of Eu and Pt \cite{Johnson1996}. XRD patterns do not show the presence of Eu-Pt alloys such as EuPt$_2$ and EuPt$_5$, but we cannot rule out the presence of these phases at the interface. 

\begin{figure}[ht]
	\centering
	\includegraphics[width=0.9\linewidth]{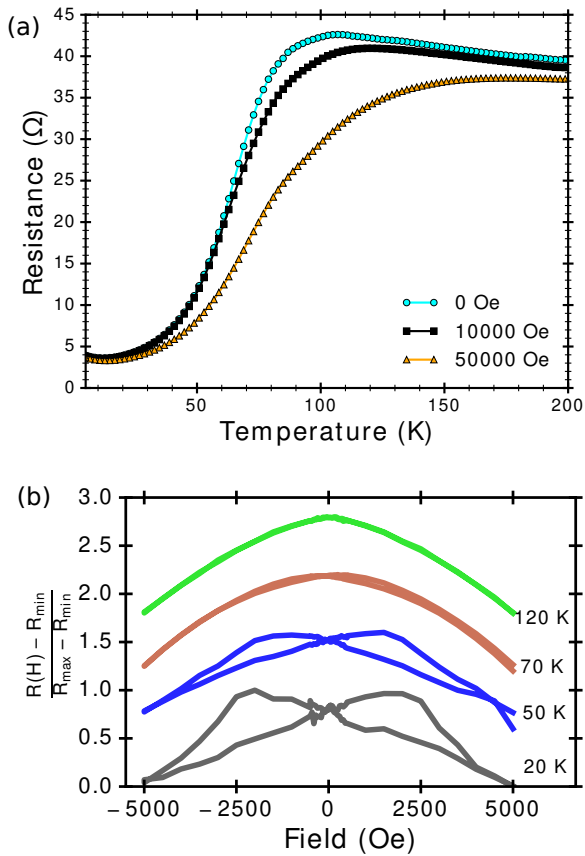}
	\caption{(a) Temperature dependence of resistance for various applied fields.  (b) Magnetoresistance curves at various temperatures. The curves are normalized and vertically shifted for clarity. R$_{max}$ and R$_{min}$ correspond to the maximum and minimum resistance for the respective curve. }
	\label{fig:fig2}
\end{figure}

The temperature dependent resistance measurements are shown in Fig. \ref{fig:fig2}. For doing transport measurements, we used optical lithography to prepare rectangular bars, with dimensions of 1 mm $\times$ 0.1 mm, and bonded gold wires directly to the film using Indium contacts.  4-probe DC measurements were performed using a Quantum Design PPMS system.  Above 70 K, the R-T curve shows an insulating behavior with a resistance of $\sim$40 $\Omega$. Since the estimated resistance for Eu-rich film is $\sim$10$^3 - 10^4 \, \Omega$  based on previous reports \cite{Konno1998,Matsumoto2004}, it is unlikely that the EuO layer is responsible for the conduction. There is also a metallic Pt layer that could contribute to the transport. The estimated resistance of the Pt overlayer at 100 K is $\sim$50 $\Omega$, based on its approximate thickness of $\sim$20 nm, device geometry and the resistivity \cite{Meyer2014}, assuming a continuous Pt film. Our measured resistance ($\sim$40 $\Omega$) suggests that at above 70 K, the Pt layer is the main conducting channel. 

The temperature response at higher temperature shows an insulting behavior, which might suggest that the conducting channel is not Pt.  It is known that granular metals have resistivity of the form $\rho \, \sim \, \exp(1/\sqrt(T))$ due to intergranular tunneling \cite{Abeles1975}, and our data are consistent with this model, showing the influence of Pt grains on the transport.  Since Pt does not wet EuO \cite{Fredrickson2016}, it is likely that our method and conditions of Pt deposition creates islands and discontinuity in the Pt layer and results in a granular film, thereby leading to the observed behavior at higher temperatures. As a matter of fact, 20 nm thick Pt films deposited on oxide substrates by various methods can be expected to be of granular nature \cite{Ryll2011,Hotovy2004,Fang2004} and lead to the observed behavior.

 The insulator-to-metal transition at $\sim$70 K is typical of doped EuO systems \cite{Oliver1972}. Above the transition temperature, the Pt layer is the primary conduction channel while the EuO is more insulating, but at the transition temperature, the EuO layer becomes more conductive and most of the current flows in this layer. 

At the lowest temperatures, the resistivity exhibits an upturn at about 17 K.  This upturn in the resistivity has been observed in older reports on doped EuO and is not well understood.  It was discussed in terms of an impurity-band hopping model and activation of additional impurity states \cite{Oliver1972,VonMolnar1970,Ohuchi2015}, or as a result of the ending of the down shift of the spin-up conduction band below the T$_c$ \cite{Sinjukow2003}.  The down shift of the spin-up conduction band toward the defect state is the possible origin of this metallic behavior.  When the down shift ends and conduction band no longer moves relative to the defect state, fewer electrons are excited to the conduction band at lower temperature and the upturn is seen.  In bulk EuO, the upturn follows activated behavior, but our sample shows a slower upturn and is also reduced in magnitude.

\begin{figure}[ht]
	\centering
	\includegraphics[width=0.9\linewidth]{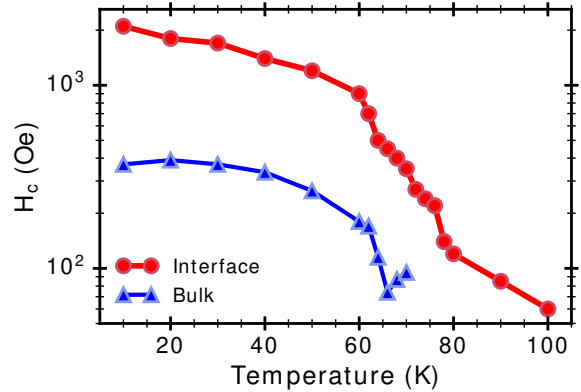}
	\caption{Temperature dependence of the MR peak positions (red circles) and coercivity obtained from the magnetic hysteresis (blue triangles).}
	\label{fig:fig3}
\end{figure}

Figure \ref{fig:fig2}(b) shows the magnetoresistance (MR) versus magnetic field plots for various temperatures.  The magnetic field is applied out of plane.  An interesting observation in the MR is the appearance and evolution of the hysteresis at various temperatures.  Usually, in ferromagnetic materials, the peaks correspond to the coercivity in the magnetic hysteresis loop.  However, the values that we have obtained from MR are very different from the values obtained by magnetometry, as can be observed by the comparison in Fig. \ref{fig:fig3}.  At 20 K, the MR peak position is close to 2000 Oe and the MR hysteresis loop remains open at a magnetic field H $>$ 4000 Oe.  The MR peak position is about a factor of 6 higher than the coercivity at the lowest measured temperature of 5 K, which suggests that the magnetic anisotropy obtained by transport is different from that obtained by bulk magnetometry measurements.

\section{Discussion}
Figure \ref{fig:fig3} suggests that the bulk magnetic properties are different from the magnetic properties probed by transport. To further investigate this, we hypothesize two scenarios that permit this observation : first is solely due to the bulk and the second is due to the Pt/EuO interface. To the best of our knowledge, no phenomena leads to such dramatic change in the coercivity values in the bulk. If the enhancement were due to the bulk, it would possibly be a result of some impure phases in the film. For the best resolutions, we were unable to detect any  contributions from impure phases in XRD or magnetic  measurements. It should be noted that the common impurities such as  Eu and Eu$_3$O$_4$ are antiferromagnetic at T$_N$ $\sim$ 90 K , 5 K resp. while Eu$_2$O$_3$ is paramagnetic. In our magnetic measurements, we failed to observe any exchange bias effects that could be caused by these impure phases that are embedded in the EuO layer. Thus, the bulk is unlikely to cause such an enhancement of the coercivity.

The second scenario in which the Pt/EuO interface leads to this enhancement seems a more plausible explanation. For this scenario, the interface would have to be a highly conductive channel, which is very plausible. According to a recent calculation \cite{Fredrickson2016}, a partially filled mid-gap state forms and is strongly localized at the Pt/EuO interface (within the first two atomic layers of EuO next to the Pt).  This state possesses primarily 4f characteristics, is spin-polarized and coincides with the bulk EuO ferromagnetic state.  This scenario also supports a large anisotropy at the interface, as discussed below.  It should be noted that this study also found the local density of states at the interface is similar to that of EuPt$_2$, suggesting the possibility that EuPt$_2$ phase may form at the interface.         

There are other possibilities for the enhancement of the anisotropy at the Pt/EuO interface.  One is due to the effect of strain at the interfaces \cite{Anuniwat2013}.  This is an unlikely scenario.  As previously discussed, EuO and Pt are unlikely to support such a strained interface because they do not wet.  Another possibility is due to interlayer exchange at the interface \cite{Leighton2000,Liu2001}, but such a scenario requires that both constituents across the interface contain elements with magnetic moments, which is not the case here.  Thus, these factors do not seem to provide satisfactory explanation for the large MR coercivity seen here, and we believe a different mechanism is at play. 

Our observations point to the effects of symmetry breaking and large spin-orbit coupling at the interface.  It was found that in Pd/Co system, the hybridization of Pd(5d) and Co(3d) states plus the large spin-orbit coupling induced by Pd leads on an enhanced perpendicular magnetic anisotropy at the interface \cite{Daalderop1994}.  Specifically, the importance of the non-zero spin-orbit energy of Pd was found to be the critical factor for turning the anisotropy of Co from in-plane to out-of-plane axis.  In our case, the Pt overlayer is expected to produce strong spin-orbit coupling.  Unlike in bulk EuO, where there is a next-nearest neighbor (NNN) coupling between Eu(4f) and Eu(5d) states, at the interface the second layer of EuO no longer has this NNN coupling with a Eu(5d) state, but has it with the Pt(5d) states across the interface \cite{Fredrickson2016}.  This Eu(4f) - Pt(5d) coupling is the primary reason for the enhanced magnetic anisotropy of this mid-gap conducting magnetic state.  In addition, EuO may also exhibit strong d-f hybridization and spin-orbit coupling \cite{Liu2012,Souza-Neto2009,Arenholz2009,vanderLaan2008}.  Such effect will be particularly strong at the surfaces or interfaces due to symmetry breaking.  The crystalline electric field at the interface, which could split the Eu(4f) states and lead to a stronger spin-orbit coupling, may result in much larger anisotropy \cite{vanderLaan2008}.

\section{Conclusions}
We have studied the magnetic and transport properties of Pt/EuO$_{1-x}$ films and found discrepancies in their bulk and interface properties.  The bulk EuO film clearly shows an enhancement of Curie temperature to about 130 K due to oxygen deficiency.  The EuO films also exhibit the expected small magnetic anisotropy as seen in the magnetic hysteresis loops.  On the other hand, a much greater magnetic anisotropy is found in the magnetotransport measurements.  The interface between EuO and Pt is found to be the source of this large anisotropy, which originates from the strong spin-orbit coupling resulting from the coupling of the Pt(5d) and Eu(4f/5d) states.  This conclusion is supported by a recent density functional theory calculation that shows strong spin-orbit coupling across the Pt/EuO interface and the presence of a 2D spin-polarized mid-gap state at the interface \cite{Fredrickson2016}.

\acknowledgements{This work was supported by the National Science Foundation (Grants No. DMR-1710512).}

\end{document}